\documentclass[floatfix,aps,twocolumn,prb,amsmath,amssymb,superscriptaddress]{revtex4-1}
\newcommand{\Eu}{EuIr$_2$P$_2$}
\newcommand{\Eui}{Eu$^{+2}$}

\usepackage{lmodern}
\usepackage{longtable}
\usepackage{bm}
\usepackage[rgb]{xcolor}
\usepackage[author={Pablo S. Cornaglia}]{pdfcomment}
	\usepackage{multirow}
	\usepackage[pdftex]{graphicx}
	\usepackage[utf8]{inputenc}

	\usepackage{tabularx}
\usepackage{array} 
\usepackage[T1]{fontenc}
\newcolumntype{C}[1]{>{\centering\arraybackslash}p{#1}}

\graphicspath{{./figs/}}

\begin{document}
\title{Magnetic properties of chiral EuIr$_2$P$_2$}
\affiliation{Centro At{\'o}mico Bariloche and Instituto Balseiro, CNEA, 8400 Bariloche, Argentina}
\affiliation{Consejo Nacional de Investigaciones Cient\'ificas y T\'ecnicas (CONICET), Argentina}

\author{D. J. Garc\'{\i}a}
\affiliation{Centro At{\'o}mico Bariloche and Instituto Balseiro, CNEA, 8400 Bariloche, Argentina}
\affiliation{Consejo Nacional de Investigaciones Cient\'ificas y T\'ecnicas (CONICET), Argentina}

\author{V. Vildosola}
\affiliation{Departamento de Materia Condensada, GIyA, CNEA (1650) San Mart\'{\i}n, Provincia de Buenos Aires, Argentina}
\affiliation{Instituto de Nanociencia y Nanotecnolog\'{\i}a CNEA-CONICET, Argentina}

\author{A. A. Aligia}
\affiliation{Centro At{\'o}mico Bariloche and Instituto Balseiro, CNEA, 8400 Bariloche, Argentina}
\affiliation{Consejo Nacional de Investigaciones Cient\'ificas y T\'ecnicas (CONICET), Argentina}
\affiliation{Instituto de Nanociencia y Nanotecnolog\'{\i}a CNEA-CONICET, Argentina}

\author{D. G. Franco}
\affiliation{Centro At{\'o}mico Bariloche and Instituto Balseiro, CNEA, 8400 Bariloche, Argentina}
\affiliation{Consejo Nacional de Investigaciones Cient\'ificas y T\'ecnicas (CONICET), Argentina}

\author{Pablo S. Cornaglia}
\affiliation{Centro At{\'o}mico Bariloche and Instituto Balseiro, CNEA, 8400 Bariloche, Argentina}
\affiliation{Consejo Nacional de Investigaciones Cient\'ificas y T\'ecnicas (CONICET), Argentina}
\affiliation{Instituto de Nanociencia y Nanotecnolog\'{\i}a CNEA-CONICET, Argentina}

\begin{abstract}
  We present a minimal model that provides a description of the magnetic and thermodynamic properties of \Eu. The model contains two exchange coupling parameters, which are calculated using Density Functional Theory, and a local easy axis magnetic anisotropy term.  The classical ground state of the system is a generalization of the well known 120$^\circ$ structure observed in triangular antiferromagnets. Monte Carlo simulations show two phase transitions as a function of the temperature. With increasing temperature, the system transitions from the ground state into a high-entropy collinear antiferromagnet, which in turn at higher temperatures presents a second order transition to a paramagnetic state. A high enough external magnetic field parallel to the anisotropy axis produces a spin-flop transition at low temperatures. The field also reduces the temperature range of stability of the collinear antiferromagnet phase and leads to a single phase transition as a function of the temperature. The reported behavior of the specific heat, the magnetization, and the magnetic susceptibility is in agreement with the available experimental data. 
  Finally, we present the magnetic phase diagrams for magnetic fields parallel 
  and perpendicular to the easy axis.
\end{abstract}


\maketitle

\section{Introduction}
A magnetic system is said to be frustrated when it is not possible to minimize all pairwise interactions simultaneously. In clean systems this can occur due to competing interactions, such as coexisting ferromagnetic and antiferromagnetic couplings. It can also have a geometrical origin of which the triangular lattice in two dimensions with nearest-neighbour antiferromagnetic interactions is the most common example. In the latter case, three nearest-neighbour magnetic moments forming a triangle in the lattice cannot minimize simultaneously their mutual interactions because it is not possible to make each magnetic moment antiparallel to the other two. For classical spins the minimal energy is obtained when the nearest-neighbour spins form a 120$^\circ$ angle\cite{diep2013frustrated}. In the presence of an Ising like magnetic anisotropy, this two dimensional system has no Curie point and no long range order even at zero temperature \cite{wannier1950antiferromagnetism}.
In the three dimensional pyrochlores such as Ho$_2$Ti$_2$O$_7$, although the interactions are predominantly ferromagnetic, geometric frustration leads to a spin-ice phase and to low energy exitations that behave as magnetic monopoles \cite{nisoli2006artificial,harris1997geometrical,borzi2016intermediate}


Another interesting frustrated material is \Eu, which has a chiral crystal structure with the \Eui\ ions forming a triangular array of helical chains \cite{lux1993kristallstrukturen}. To the best of our knowledge \Eu\ is the only intermetallic magnetic compound known to crystallize in the chiral P3$_2$21 trigonal space group.
It presents two phase transitions as a function of the temperature which have been associated with antiferromagnetic orderings of local magnetic moments at the \Eui\ ions \cite{franco2021synthesis}. The interest in \Eu\ is fueled by its chiral structure and signatures of magnetic frustration that make it a likely candidate for the observation of exotic magnetic textures\cite{gao2020fractional,muhlbauer2009skyrmion,yu2010real,karube2018disordered}. 

Magnetic frustration in \Eu\ is due to both competing interactions (the possitive Curie-Weiss temperature indicates predominant ferromagnetic interactions) and geometric frustration due to its triangular structure.
Additionally, the lack of inversion symmetry due to the chiral crystal structure of \Eu\ allows a Dzyaloshinskii–Moriya interaction (DMI) term, which can stabilize nontrivial spin textures as skyrmions\cite{rosales2015three,villalba2019field}. 
 For simplicity we neglect the DMI in this work. 




We present a detailed study of the magnetic properties of \Eu\ to determine the nature of the observed antiferromagnetic phases. A magnetic field-temperature phase diagram is obtained using Monte Carlo simulations, 
as done previuously for similar systems \cite{miya85,wata01,miya10}. 
To that aim we construct a Hamiltonian for the magnetic moments with coupling parameters estimated using {\it ab initio} calculations. In addition 
to the usual high-temperature paramagnetic phase, we find a non-collinear antiferromagnetic ground state, which is a generalization of the 120$^\circ$ two-dimensional state, and an intermediate temperature high entropy collinear antiferromagnetic state. This leads to two phase transitions as a function of temperature (for low enough external magnetic fields) and a two peak structure in the specific heat, in agreement with the available experimental data. 

The rest of this paper is organized as follows: 
In Sec. \ref{sec:model} we present the 
Hamiltonian for the magnetic moments of \Eu\ and determine its magnetic coupling constants through Density Functional Theory (DFT) calculations. In Section \ref{sec:magprop} we determine the classical ground state of the magnetic Hamiltonian and analyze the effect of a magnetic anisotropy term. We also present Monte Carlo simulations of the thermodynamic properties. Finally, in Sec. \ref{sec:concl} we summarize our main results and conclusions.

\section{Model for the magnetic interactions} \label{sec:model}
In this Section we analyze the magnetic structure of \Eu. We propose a simple Hamiltonian to describe its magnetic properties and perform DFT calculations to determine the model parameters.

\subsection{Crystal structure}

\begin{figure}[t]
    \begin{center}
      \includegraphics[width=0.4\textwidth]{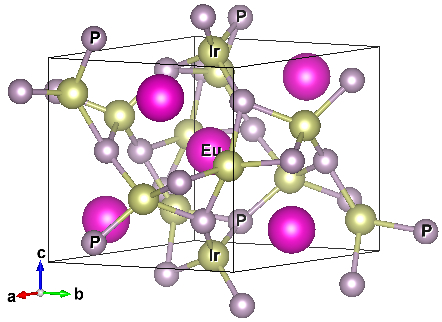}
    \end{center}
    \caption{(Color online) Crystal structure of \Eu. The thin solid lines indicate the unit cell containing three Eu atoms.}
    \label{fig:crystalstruc}
\end{figure}

The space group of \Eu\ is the non-symmorphic one P3$_2$21, belonging to the trigonal crystal system (see Fig. \ref{fig:crystalstruc}). The basis vectors
${\bf a}$ and ${\bf b}$ have the same magnitude and form an angle of 
$120^\circ$ between them. The third basis vector ${\bf c}$ is perpendicular
to the other two. 
The lattice parameters are $a=6.72 \AA$ and $c=7.12 \AA$.
While the rotation in $120^\circ$ ($C_3$) and the translation
in ${\bf c}/3$ ($t_{{\bf c}/3}$) are not symmetry operations of the system,
the product $C^{2}_3 t_{{\bf c}/3}$ is a screw axis operation that leaves 
the system invariant.

The lattice of \Eui\ ions is similar to the ABC stacking of hexagonal layers in the face centered cubic lattice. The successive layers are however shifted away from the high-symmetry point of the adjacent layers (see Fig \ref{fig:Eucrystal}).  

There are three Eu atoms in the unit cell. Their positions are
\begin{align}\label{pos}
\text{ Eu$_1$:\quad  }& {\bf r}_1=0.604{\bf a} + {\bf c}/6 \nonumber \\
\text{ Eu$_2$:\quad  }& {\bf r}_2=0.396{\bf a} + 0.396{\bf b} +{\bf c}/2 \nonumber \\
\text{ Eu$_3$:\quad  }& {\bf r}_3=0.604{\bf b} + 5{\bf c}/6 
\end{align}%
Noting that 
$C^{2}_3{\bf a}=C_3{\bf b}=-{\bf a}-{\bf b}$, $C^{2}_3{\bf b}={\bf a}$, 
it is easy to see that the screw axis operation $C^{2}_3t_{{\bf c}/3}$ permutes
the Eu atoms (changing also the unit cell) as 
Eu$_1$ $\rightarrow$ Eu$_2$ $\rightarrow$ Eu$_3$ $\rightarrow$ Eu$_1$.

The Eu atoms are expected to be in the \Eui\ electronic configuration with 7 electrons on the 4f orbital. Following Hund's rules, the total spin per \Eui\ ion is expected to be $S=7/2$, the angular momentum $L=0$, and  the total angular momentum $J=7/2$. Since the hybridization of the Eu 4f orbital is negligible, a local magnetic moment $\mu=g\mu_B J$ with $g=2$ is expected at each \Eui\ ion. 
However, a small admixture of states with $L=1$ is also expected,
which can slightly modify $g$ \cite{betancourth2019magnetostriction}.

\subsection{Hamiltonian}

\begin{figure}[t]
    \begin{center}
       \includegraphics[width=0.4\textwidth]{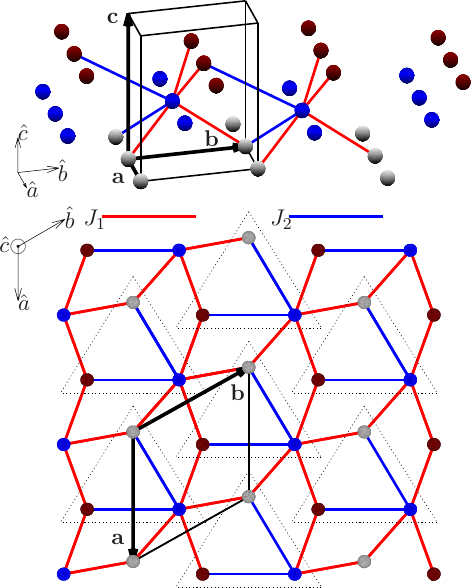}
    \end{center}
    \caption{(Color online) Crystal structure of the \Eui\ ions in the EuIr$_2$P$_2$ compound.
      The unit cell containing three \Eui\ ions (Eu$_1$, Eu$_2$, and Eu$_3$ are represented using different colors) is indicated using thin solid lines. The first (second) nearest neighbours of the Eu$_2$ atoms (blue symbols) are indicated using red (blue) lines.
  The \Eui\ ions inside a dotted line triangle belong to the same one dimensional spiral chain of antiferromagnetically coupled magnetic moments (see text).}
    \label{fig:Eucrystal}
\end{figure}

{\it Ab initio} calculations  (for details see Appendix \ref{ap:couplings}) indicate an insulating state with a band gap $\Delta\sim 0.4$eV. 
Short range exchange interactions (generated by a superexchange mechanism) are therefore expected between the magnetic moments on the Eu$^{2+}$ ions\footnote{We do not consider here dipolar interactions which, as we discuss below, contribute to the magnetic anisotropy.}.
We model these magnetic interactions using the following effective Hamiltonian that considers exchange couplings up to second neighbour magnetic moments:
\begin{equation}
  \mathcal{H}_{m}= -\sum_{\left< i,j\right>} J_{1} \mathcal{J}_i \cdot {\mathcal J}_j - \sum_{\left<\left< i,j\right>\right>} J_{2} \mathcal{J}_i\cdot \mathcal{J}_j,
    \label{eq:magham}
\end{equation}
where $\mathcal{J}_i$ is the total angular momentum at site $i$,
and $J_{1}$ ($J_2$) is the exchange coupling between nearest-neighbour (next-nearest-neighbour) magnetic moments (see Fig. \ref{fig:Eucrystal}).
We use below units such that $k_B$=$g\mu_B$=1. 
Because of the large magnitude of $|\mathcal{J}_i|=J=7/2$, we will
treat $\mathcal{J}_i$ as a classical vector.

We obtain the coupling constants of $ \mathcal{H}_{m}$ through {\it ab initio} calculations of the total energy of the system with the local magnetic moments fixed in different configurations. The details of these calculations are presented in Appendix \ref{ap:couplings}. We obtain a ferromagnetic (FM) first nearest-neighbour interaction ($J_1=0.28$ K$>0$) and an antiferromagnetic (AF) second nearest-neighbour interaction ($J_2=-0.45$ K$<0$).


\section{Magnetic properties} \label{sec:magprop}
\subsection{Ground State}
Each \Eui\ has four nearest neighbours (NN), two on each one of the adjacent layers along the $\hat{c}$-axis. For example Eu$_2$ at ${\bf r}_2$ [see Eqs. (\ref{pos})] has two NN Eu$_1$ atoms lying at ${\bf r}_1$ and ${\bf r}_1+{\bf b}$ and two NN Eu$_3$ atoms at 
${\bf r}_3$ and ${\bf r}_3+{\bf a}$. The remaining NN positions
can be obtained using screw axis and translation symmetries.

Each \Eui\ has also two next-nearest neighbours (NNN), one on each 
of the adjacent layers along the $\hat{c}$-axis (see Fig. \ref{fig:Eucrystal}).
The NNN distance (5.17 \AA) is similar to the NN one (4.25 \AA).
The Eu$_2$ site at ${\bf r}_2$ has a NNN Eu$_1$ at position 
${\bf r}_1^{\prime}={\bf r}_1-{\bf a}$ and a NNN Eu$_3$ at position 
${\bf r}_3^{\prime}={\bf r}_3-{\bf b}$. Furthermore, it is easy to see 
that under the screw axis operation $C^{2}_3t_{{\bf c}/3}$, these 
positions transform as 
${\bf r}_1^{\prime} \rightarrow {\bf r}_2 \rightarrow {\bf r}_3^{\prime} \rightarrow {\bf r}_1^{\prime}+{\bf c}$.
Therefore, if the NN coupling $J_1$ is neglected, 
retaining only the dominant NNN coupling $J_2$, 
the system can be seen as formed by one-dimensional (1D) spiral chains along the $\hat{c}$ axis with antiferromagnetic (AF) intrachain couplings $J_2$. 
To be more specific, the spin at ${\bf r}_1^{\prime}$ can point in any
direction ${\bf n}$ and the coupling $J_2$ is optimized taking
$\mathcal{J}_{{\bf r}_2} \parallel - {\bf n}$, 
$\mathcal{J}_{{\bf r}_3^{\prime}} \parallel {\bf n}$,
$\mathcal{J}_{{\bf r}_1^{\prime}+{\bf c}} \parallel -{\bf n}$,
and so on,  so that the magnetic unit cell is doubled in the
$c$ direction, with the moments at positions differing in
${\bf c}$ pointing in opposite directions.

There are three different of these 1D chains per unit cell related 
by the screw axis operation $C^{2}_3t_{{\bf c}/3}$ [see below Eq. (\ref{pos})]. Each of these 1D chains form a triangular lattice 
in the $a$--$b$ plane, with basis vectors ${\bf a}$ and ${\bf b}$. 

Including the ferromagnetic (FM) NN couplings $J_1$, one realizes that two Eu sites at a distance $a$ are connected by two paths, each one involving an NNN AF and one NN FM coupling. 
This implies an effective AF coupling between 1D chains at 
a distance of one lattice parameter $a$.
Therefore the system is frustrated as the
simple two-dimensional (2D) triangular lattice with AF NN couplings \cite{collins1997review,kawamura1984phase,diep2013frustrated}.
With this picture in mind, to obtain the ground state (GS) configuration, assuming a classical magnetic moment description and for large enough AF $J_2$, we can view the system as an effective 2D triangular lattice (each site corresponds to an AF 1D spiral chain). The classical GS configuration of a triangular lattice with nearest-neighbor AF couplings is the well known 120$^\circ$ structure, where the magnetic moments rotate $120^\circ$ from one site to the next in a given $a$--$b$ plane. The magnetic unit cell increases by a factor 
three and the basic vectors become ${\bf a}-{\bf b}$ and 
$2{\bf a}+{\bf b}$. 

To analyze the order among the three 1D chains inside the non-magnetic
unit cell, we note that for example, as mentioned at the beginning of this section, Eu$_2$ at ${\bf r}_2$ has two NN Eu$_1$ atoms lying at 
${\bf r}_1$ and ${\bf r}_1+{\bf b}$. 
The magnetic moments of these Eu$_1$ atoms form an angle
of 120$^\circ$ according to the above discussion. Then to optimize 
the NN FM coupling $J_1$, it is convenient that the magnetic moment 
of the Eu$_2$ at ${\bf r}_2$
has the direction of the sum of the two moments of the Eu$_1$ atoms, 
forming an angle of 60$^\circ$ with each of them. This completes the 
description of this phase that we call AF120. 
The magnetic unit cell contains 18 Eu atoms with an energy per \Eui\ ion given by:

\begin{equation}
  E_{AF120}= J^2 (J_2-J_1)
  \label{GSEnoD}
\end{equation}
\begin{figure}[h]
  \centering
   \includegraphics[width=0.3\textwidth]{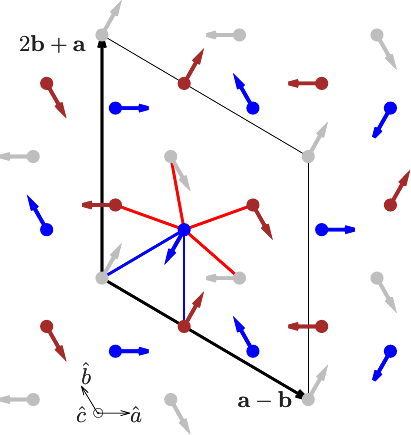}
   \caption{(Color online) Magnetic configuration of a unit cell layer of the AF120 ($\perp\hat{c}$) state. The orientation of the magnetic moments on the Eu ions is indicated using arrows. The shape of the magnetic unit cell in the $a$--$b$ plane is indicated with solid lines. The color coding is as in Fig. \ref{fig:Eucrystal}.} 
  \label{fig:munitcell}
\end{figure}
Due to the invariance of the Hamiltonian (\ref{eq:magham}) under a global SU(2) rotation of the magnetic moments, we can construct the AF120 state with all the magnetic moments perpendicular to the 
$\hat{c}$-axis. The resulting GS configuration is shown in Fig. \ref{fig:munitcell}.

The AF120 magnetic moment configuration is the ground state if the NNN coupling is antiferromagnetic ($J_2<0$) and the NN coupling is ferromagnetic ($J_1>0$) and smaller than $-2J_2$. For  $J_1>-2J_2$ the ground state is ferromagnetic, while for $J_1<0$ it is a type A antiferromagnet (the magnetic moments are ferromagnetic on each of the hexagonal layers stacked along the $\hat{c}$-axis and the sign of the orientations is opposite to the one on the nearest neighbouring layers).

To complete the description of the magnetic properties we turn on an effective magnetic anisotropy term:
 \begin{equation}
   \mathcal{H}_D=   -D\sum_i (\mathcal{J}_{i}\cdot \hat{c})^2 
   \label{eq:anisotropy}
 \end{equation}
 There are two main sources of magnetic anisotropy: the crystal field induced coupling of the $L=0$ multiplet to higher $L$ multiplets and the dipolar interaction \cite{betancourth2019magnetostriction}. The available experimental results for this system show a significant anisotropy in the magnetic susceptibility \cite{franco2021synthesis}. As we discuss below, these properties are consistent with an easy axis anisotropy ($D>0$).  
 {\it Ab initio} calculations for \Eu\ indicate a crystal field induced easy axis contribution to the anisotropy parameter $D$ when the spin-orbit coupling is turned on (see Appendix \ref{ap:couplings}).  For completeness
  we analyze the effect of $D$ for a wide range of values.

  For an easy plane anisotropy ($D<0$), the GS order and energy are not altered and all magnetic moments are perpendicular to the $\hat{c}$-axis (see Fig. \ref{fig:munitcell}).

\begin{figure}[h]
  \centering
  \includegraphics[width=0.5\textwidth]{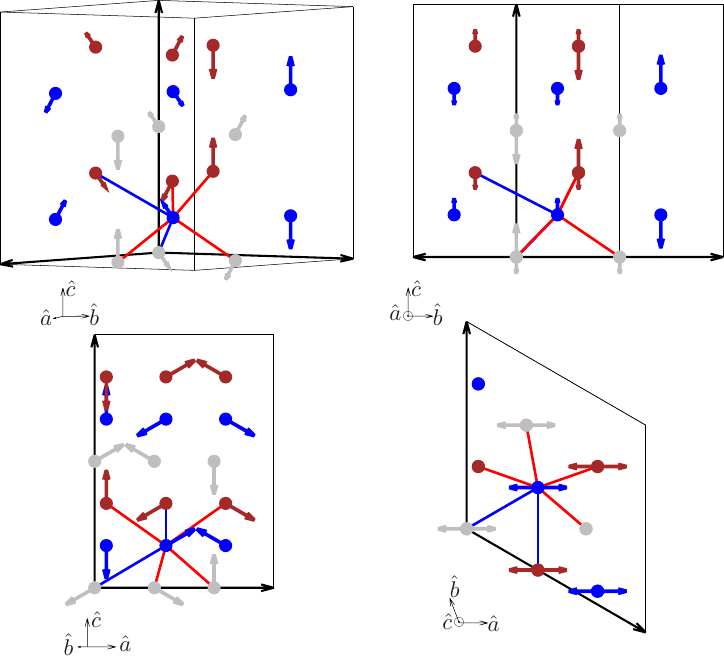}
  \caption{(Color online) Magnetic unit cell of the AF120 state. The magnetic moments are contained in $a$--$c$ planes and one out of three magnetic moments is parallel or antiparallel to the $\hat{c}$-axis. The color coding is as in Fig. \ref{fig:Eucrystal}. } 
  \label{fig:AFD0}
\end{figure}

For an easy axis anisotropy along $\hat{c}$ ($D>0$) it is energetically favorable to increase the projection of the magnetic moments along the $\hat{c}$-axis. The resulting antiferromagnetic order (which we name AFD) can be constructed from an AF120 order in the $a$--$b$ plane (see Fig. \ref{fig:munitcell}) rotating all magnetic moments by $90^\circ$ around the $\hat{c}\times\hat{a}$-axis. Doing so, one out of three magnetic moments is parallel to the $\hat{c}$-axis, while the other two form a $\pi/3$ (or $2\pi/3$) angle with it. This is presented in Fig. \ref{fig:AFD0}, where we have used the rotational symmetry around the $\hat{c}$-axis of the Hamiltonian to put the magnetic moments in the $a$-$c$ plane. For $D>0$ it is energetically favorable to decrease the $\pi/3$ angle (or increase the $2\pi/3$ angle) by $\delta\theta>0$ to enhance  the projection of the magnetic moments along the $\hat{c}$-axis (see Fig. \ref{fig:AFD}). A three dimensional animation of the magnetic unit cell in the AFD state is available in the Supplemental Material \footnote{See Supplemental material at URL for the following: an
animation of the magnetic unit cell in the AFD state.}.

\begin{figure}[h]
  \centering
  \includegraphics[width=0.15\textwidth]{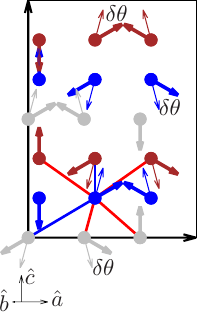}
  \caption{(Color online)  Magnetic unit cell of the AF120 (thick style arrows) and the AFD (thin style arrows) states. The magnetic anisotropy term produces a tilting (by an angle $\delta \theta$) of two out of three magnetic moments that increases the absolute value of their projection along the $\hat{c}$-axis. The color coding is as in Fig. \ref{fig:Eucrystal}. }
  \label{fig:AFD}
\end{figure}

\begin{figure}[t]
    \begin{center}
       \includegraphics[width=0.45\textwidth]{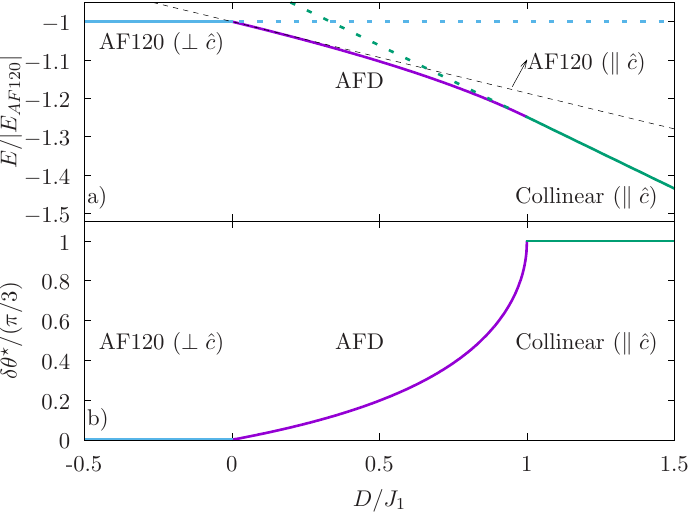}
    \end{center}
\caption{(Color online) a) Energy, for $J_2=-1.68J_1$, of the different magnetic moment configurations considered. For negative $D$ the ground state is the AF120 configuration with all magnetic moments perpendicular to the $\hat{c}$-axis, while for $D> J_1$ it is an antiferromagnetic state with all magnetic moments parallel to the $\hat{c}$-axis.  b) Shift angle $\delta\theta^\star$ of the ground state configuration as a function of the magnetic anisotropy $D$. }
    \label{fig:phasesD}
\end{figure}

The shift $\delta\theta$ does not change the angle between next-nearest-neighbour magnetic moments, which remain antiferromagnetically ordered. However, the nearest neighbor interaction is modified by a non-zero $\delta \theta$. In the AF120 configuration, each \Eui\ ion magnetic moment has three possible orientations. As a consequence, twelve first neighbour couplings need to be analyzed to determine how the interaction energy is modified by $\delta \theta$.  Eight of the nearest neighbour interactions see a decrease of the relative angle between magnetic moments by $\delta\theta$ while the remaining four see an increase by $2\delta \theta$.
The shift in the angle that minimizes the energy therefore results from the competition between the anisotropy and the nearest neighbour interaction $J_1$. 
The magnetic energy per \Eui\ ion is given by:
\begin{align}
  E(\delta \theta) =&  J^2 \left[-\tfrac{2}{3} D \cos^2\left(\tfrac{\pi}{3}-\delta\theta \right)-\tfrac{D}{3}
   -\tfrac{4}{3} J_1 \cos \left(\tfrac{\pi }{3}-\delta \theta \right) \right.\nonumber\\ & - \left. \tfrac{2}{3} J_1 \cos \left(\tfrac{\pi }{3} + 2\delta\theta \right)+ J_2\right] \label{eq:DJ1}
\end{align}
Minimizing this energy w.r.t. $\delta\theta$ leads to
\begin{equation}
  \tan \delta \theta^\star=\frac{3  J_1-\sqrt{3} \sqrt{(3 J_1-D) (J_1-D)}}{ \sqrt{3} J_1+3 \sqrt{(3 J_1-D) (J_1-D)}}
\end{equation}
and
\begin{equation}\label{eq:EAFD}
  E_{AFD}= J^2\left(J_2-\frac{D^2-6 J_1^2}{3 (D-2 J_1)}\right)
\end{equation}
As $D$ increases, $\delta\theta^\star$ increases and reaches $\pi/3$ for $D=J_1$, where the magnetic moments are collinear. For larger values of $D$ this is the lowest energy configuration. 
If $D\geq J_1$ the GS energy is given by
\begin{equation}\label{eq:Ecoll}
	E_{coll}= J^2\left(J_2-D-\frac{2 J_1}{3}\right)
\end{equation}
These results are summarized in the phase diagram of Fig. \ref{fig:phasesD}.

We focus below on the $0<D<J_1$ parameter regime which appears to be the experimentally relevant situation and consistent with the parameters estimated using DFT calculations.
Although the energy of the collinear state is larger than that of the AFD one for this range of magnetic anisotropy, the collinear state has a high degeneracy which can make its free energy lower than the AFD state at finite temperatures. The high degeneracy of the collinear state can be understood using again an effective triangular lattice to describe the system. The collinear (Ising) state in the antiferromagnetic triangular lattice has a high configurational degeneracy due to frustration (see e.g. Ref. \onlinecite{wannier1950antiferromagnetism}).

We consider below the following Zeeman coupling of the magnetic moments to an external magnetic field  ${\bf{B}}$: 
\begin{equation}
  H_Z = -g\mu_B\sum_i {\mathcal{J}_i}\cdot {\bf B}.
\end{equation}

\subsubsection{External magnetic field parallel to the easy axis ($\mathbf{B}\parallel \hat{c}$)}
\begin{figure}[h]
  \centering
\begin{tabular}{p{0.25\textwidth} p{0.25\textwidth}}
  \vspace{0pt} \includegraphics[width=0.3\textwidth]{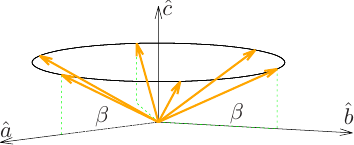}
\end{tabular}
 \caption{(Color online) Schematic representation of the relative orientation of the magnetic moments in the \Eui\ ions. The magnetic field is parallel to the $\hat{c}$-axis and produces an increase of the projection of the magnetic moments along the same axis. } 
  \label{fig:angles}
\end{figure}

For a large enough magnetic field applied parallel to the $\hat{c}$-axis (${\bf B}=B\hat{c}$) there is a spin-flop transition. The spin-flop state can be obtained starting from the AF120 state with all magnetic moments in the $a$--$b$ plane and rotating them by an angle $\beta$ (see Fig. \ref{fig:angles}) preserving the 120$^\circ$ angle between the projections of the magnetic moments on the $a$--$b$ plane.
Specifically, taking the AF120 state described in detail at the 
beginning of Section III A, in which the 18 magnetic moments 
$\mathcal{J}_i$ lie in the $a$--$b$ plane, each of them 
is rotated an angle $\beta$ around the axis 
$\hat{c}\times \mathcal{J}_i$.

The projection of the magnetic moments along the $\hat{c}$-axis for a tilting angle $\beta$ is $J\sin(\beta)$
while the $a$--$b$ plane projection is reduced to $J\cos(\beta)$. This leads to an energy (per \Eui\ ion)
\begin{align}
  E(\beta)=& J^2 \left[-D \sin ^2(\beta )-2 J_1 \sin ^2(\beta )- J_1 \cos ^2(\beta)\right.\nonumber \\ &\left. + J_2 \cos (2 \beta )\right]-B J \sin (\beta ) \label{Eflop}
\end{align}

The angle that minimizes the energy of this state for a given magnetic field [$|B|<B^\parallel_{pol}=-2 J(D+J_1+2J_2)$] is 
\begin{equation}
  \tan \beta^\star=\left(\frac{B}{\sqrt{4 J^2 (D+ J_1+2 J_2)^2-B^2}}\right),
\end{equation}
and the corresponding energy: 
\begin{equation} \label{eq:eflop}
  E^\parallel_{flop}=\frac{B^2}{4 (D+ J_1+2 J_2)}+ J^2 (J_2-J_1).
\end{equation}
For $B>B_{flop}$ where

\begin{equation}\label{eq:flopafd}
  B_{flop}\gtrsim 2 J\sqrt{\frac{D(3 J_1-D)(-D-J_1-2J_2)}{6 J_1- 3 D}}
\end{equation}
the spin-flop state becomes the ground state of the system (see Appendix \ref{ap:AFDB}).
However, it is known that this state has a low entropy and 
becomes unstable as the temperature is increased \cite{miya85,wata01,miya10}.

For $|B|>B^\parallel_{pol}$, the lowest energy configuration is fully polarized ($\beta^\star=\pi/2$) and has an energy
\begin{equation}
  E_{pol}=-J^2(J_2+2J_1+D)-J B.
  \label{eq:ferro}
\end{equation}

The magnetic field that makes the energies of the spin-flop and the collinear states equal (this is relevant at finite temperatures where the collinear state is favored by entropic effects) is given by:
\begin{equation}
  B_{flop2}=\frac{2 J}{\sqrt{3}}\sqrt{(3 D- J_1)(-D-J_1-2J_2)}.
\end{equation}

\subsubsection{External magnetic field perpendicular to the easy axis ($\mathbf{B}\perp \hat{c}$)}
For a magnetic field applied perpendicular to the $\hat{c}$-axis, the ground state can be constructed starting from the AFD configuration in a plane perpendicular to the magnetic field, and tilting the magnetic moments in the direction of the magnetic field by an angle $\gamma$. The ground state configuration is similar to the spin-flop state and no spin-flop transition is expected in this case with increasing magnetic field. At high enough magnetic fields, however, the tilting angle reaches $\pi/2$ and the ground state is fully polarized.

The energy per \Eui\ ion in the state AFD as a function of the tilting angle $\gamma$ reads
\begin{equation}
  E_{AFD}(\gamma)= E_{AFD} \cos^2(\gamma)+E_1\sin^2(\gamma) -B J \sin (\gamma).
  \label{eq:enperp}
\end{equation}
where $E_1=-J^2(J_2+2J_1)$ is the energy per \Eui\ ion of a state fully polarized in a direction perpendicular to the easy axis. Minimizing w.r.t. $\gamma$ leads to
\begin{equation}
  \tan(\gamma^\star)=\frac{J B}{\sqrt{4 (E_{AFD}-E_1)^2- J^2B^2 }}
  \label{eq:tilbperp}
\end{equation}
and 
\begin{equation}
  E^\perp_{AFD}(B)=E_{AFD}-\frac{3 J ^2 B^2}{4 E_{AFD}-4 E_1},
  \label{eq:Etilbperp}
\end{equation}
for fields below the polarization field $B_{pol}^\perp = 2 (E_{AFD}-E_1)/J$,
which sets the threshold value for a fully polarized state with energy:
\begin{equation}
  E^{\perp}_{pol}=E_1-JB.
\end{equation}

\subsection{Finite temperatures} \label{sec:MF}
We perform classical Monte-Carlo simulations using the ALPS code library \cite{bauer2011alps,ALBUQUERQUE20071187} to calculate the specific heat, the magnetization, and the magnetic susceptibility. We present results for systems 
with $24\times24\times24$ crystal unit cells, but we analyzed smaller systems to rule out significant finite size effects.  
 \begin{figure}[t]
\begin{center}
       \includegraphics[width=0.45\textwidth]{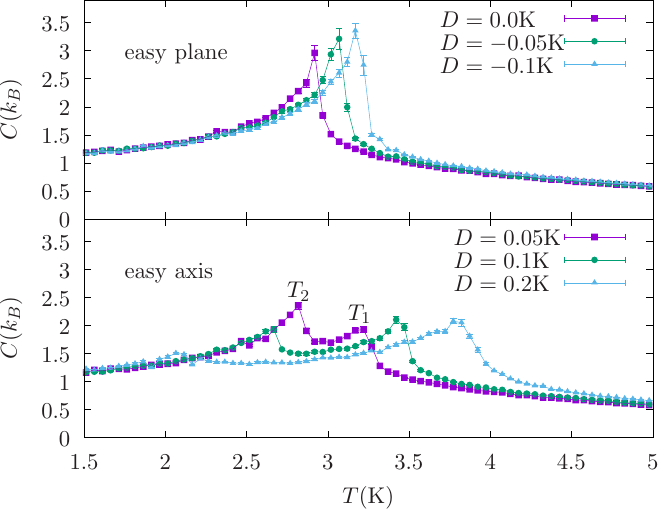}
    \end{center}
    \caption{(Color online) Specific heat as a function of the temperature for different values of the magnetic anisotropy $D$ and $B=0$. The interaction parameters are $J_1=0.28$K and $J_2=-0.47$K.}
    \label{fig:CvsT_D}
\end{figure}

Figure \ref{fig:CvsT_D} presents the magnetic contribution to the specific heat per \Eui\ ion as a function of the temperature for different values of the magnetic anisotropy coefficient $D$. For an easy plane anisotropy $D<0$, a single peak in the specific heat is obtained. It signals the transition from a high-temperature paramagnetic state to a low-temperature antiferromagnetic AF120 state with the magnetic moments lying on the $a$--$b$ plane. Increasing the absolute value of $D$ makes the AF120 more stable and suppresses the fluctuations of the magnetic moments away from the $a$--$b$ plane, which leads to an increase in the N\'eel temperature. An easy axis anisotropy ($D>0$) has a qualitatively different effect in the specific heat (see lower panel in Fig. \ref{fig:CvsT_D}). It splits the peak in the specific heat into high-temperature $T_1$ and low-temperature $T_2$ peaks.
These two peaks mark two phase transitions. A high-temperature transition from a paramagnetic state to a collinear high-entropy antiferromagnetic state at $T_1$, and a low-temperature transition from this state to the AFD state at $T_2$. The high entropy of the collinear state $S_{coll}$ reduces its free energy $F_{coll}=E_{coll}-T S_{coll}$ below the one of the AFD state $F_{AFD}=E_{AFD}-T S_{AFD}$ in the range of temperatures $[T_2,T_1]$.
The reported specific heat for this compound presents two peaks at $\sim 3$K and $\sim 5$K \cite{franco2021synthesis}, which is consistent  with an easy axis anisotropy. 

\subsubsection{External magnetic field parallel to the easy axis ($\mathbf{B}\parallel \hat{c}$)}
\begin{figure}[t]
    \begin{center}
       \includegraphics[width=0.5\textwidth]{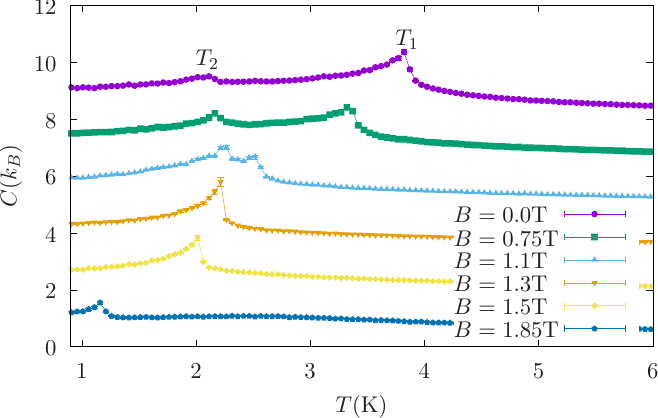}
    \end{center}
    \caption{(Color online) Specific heat as a function of the temperature for different values of the external magnetic field $B$ along the $\hat{c}$-axis (the curves are shifted by 1.6$k_B$). The anisotropy parameter is $D=0.2$K. Other parameters as in Fig. \ref{fig:CvsT_D}. }
    \label{fig:CvsT_B}
\end{figure}

In what follows we analyze the effects of an external magnetic field on the thermodynamic properties of the system for an easy axis anisotropy term $D=0.2$K.
The specific heat (per \Eui\ ion) is presented in Fig. \ref{fig:CvsT_B} as a function of the temperature for different values of the external magnetic field parallel to the $\hat{c}$-axis.
As expected, the N\'eel transition temperature $T_1$ to a collinear antiferromagnet is reduced monotonically by an increasing magnetic field. The transition temperature $T_2$ increases slowly with increasing magnetic field. This is caused by a reduction of the energy of the AFD state with respect to the collinear state as the magnetic field is increased. While the energy of the collinear state is not modified by the external magnetic field, the energy of the AFD state is reduced by it (see Appendix \ref{ap:AFDB})

For fields larger than $\sim 1.25$T the two peaks in the specific heat merge into a single transition. For fields larger than $B_{flop}= 1.2$T and smaller that $B_{pol}\sim 2.4$T, the transition at low temperature is to an antiferromagnetic spin-flop state which is the ground state of the system for those fields. For $B>B_{pol}\simeq 2.4$T, the ground state of the system is  fully polarized but no clear sign of a paramagnetic-ferromagnetic transition is observed as a function of the temperature. The magnetization increases monotonically with decreasing temperature and no jumps or kinks are observed.

\begin{figure}[t]
    \begin{center}
       \includegraphics[width=0.5\textwidth]{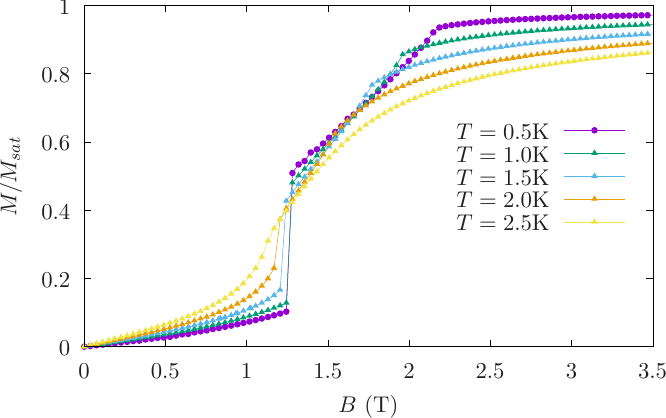}
    \end{center}
    \caption{(Color online) Magnetization as a function of the external magnetic field (parallel to the easy axis $\hat{c}$) for different values of temperature. Other parameters as in Fig. \ref{fig:CvsT_B}.}
    \label{fig:Mag_para}
\end{figure}

Figure \ref{fig:Mag_para} presents the magnetization as a function of the magnetic field for different values of the temperature. Increasing the magnetic field from zero in the low-temperature regime $T<T_2$,  the system is in the AFD state with a small magnetization which increases with increasing magnetic field. For fields $B>B_{flop}$ the energy of the AFD state is larger than the energy of the spin-flop state (their entropies are expected to be similar). This leads to a spin-flop transition to a state with a larger magnetization. As the magnetic field is further increased, the magnetization increases linearly, as the magnetic moments tilt increasingly in the direction of the field. At a field $B\sim 2$T there is a kink in the magnetization that signals the transition from the spin-flop state to a paramagnetic state highly polarized by the large magnetic field. This kink coincides with the peak in the specific heat used to determine $T_1$. As the temperature is increased approaching $T_1$ the jump in the magnetization at the spin-flop transition decreases and vanishes at $T=T_1=T_2$ where the two peaks in the specific heat merge.
For temperatures $T_2<T<T_1$ the zero field state is a collinear antiferromagnet. No jump in the magnetization is observed in this case for $B=B_{flop2}\sim 1.15$T where the energy of the spin-flop state is equal to the energy of the collinear state. The absence of a spin-flop transition is due to entropic effects, as the large entropy of the collinear state makes its free energy lower. We can estimate the field at which the spin-flop transition is expected to occur: at zero field and at the temperature $T_2$ the free energy of the collinear and AFD states are equal which means that:
\begin{equation}
  E_{AFD}-E_{coll}=T_2(S_{AFD}-S_{coll}),
\end{equation}
where $S_\alpha$ is the entropy of state $\alpha$.
The free energies of the spin-flop state and the collinear state  at a temperature $T_2$ would be equal for a field $B^\star$ 
\begin{equation}
  E^\parallel_{flop}(B^\star)-E^\parallel_{coll}=T_2(S_{flop}-S_{coll})
\end{equation}
Assuming $S_{AFD}\sim S_{flop}$ we obtain $E^\parallel_{flop}(B^\star)-E_{coll}\sim 0.15 $K. Using Eq. (\ref{eq:eflop}) and Eq. (\ref{eq:Ecoll}), we obtain $B^\star\sim 1.22$T.  For fields $B\gtrsim B^\star$, the system is however already in the paramagnetic phase. As a consequence, entropic effects preclude the spin-flop transition from happening out of the collinear state by increasing the magnetic field.

The most salient feature in the magnetization as a function of the magnetic field (for $T_2<T<T_1$) is a maximum in the slope which coincides with the transition from collinear AF to paramagnet identified using the specific heat.

\begin{figure}[t]
\begin{center}
       \includegraphics[width=0.45\textwidth]{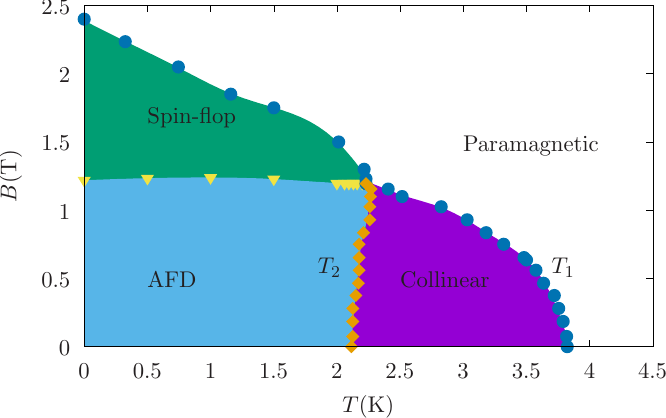}
    \end{center}
    \caption{Phase diagram for an external magnetic field parallel to the easy axis $\hat{c}$. Other parameters as in Fig. \ref{fig:CvsT_B}.}
    \label{fig:PD}
\end{figure}

The transitions lines obtained from an analysis of the ground state energies and from the specific heat and magnetization data are presented in a phase diagram $B$ vs $T$ of Fig. \ref{fig:PD}. A similar phase diagram has been obtained in ABX$_3$ compounds were A is an alkali metal, B is a transition metal, and X is an halogen atom\cite{collins1997review,diep2013frustrated}. 

\subsubsection{External magnetic field perpendicular to the easy axis ($\mathbf{B}\perp \hat{c}$)}
\begin{figure}[t]
    \begin{center}
       \includegraphics[width=0.5\textwidth]{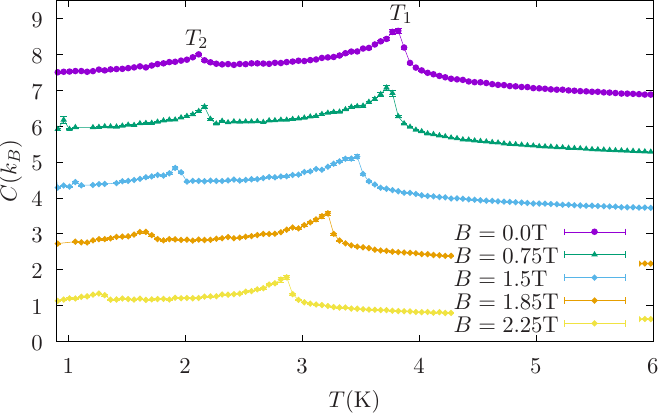}
    \end{center}
    \caption{(Color online) Specific heat as a function of the temperature for different values of the external magnetic field $B$ along the $\hat{a}$-axis (the curves are shifted by 1.6$k_B$). Other parameters as in Fig. \ref{fig:CvsT_B}.}
    \label{fig:CvsT_Brot}
\end{figure}

The behavior of the system when the magnetic field is applied perpendicular to the easy axis differs significantly from the parallel configuration. In this case, the magnetic moments tilt in the direction of the magnetic field both in the AFD and in the collinear AF phases.

The specific heat as a function of the temperature (see Fig. \ref{fig:CvsT_Brot}) shows a slower decrease of the position of the 
high-temperature peak $T_1$ with increasing magnetic field compared to the parallel case. This is expected because, contrary to the parallel situation, a perpendicular field allows a significant reduction of the energies of the AF phases  by tilting the magnetic moments in the direction of the magnetic field. 
For low fields ($B<1$T) the position of the low-temperature peak $T_2$, remains approximately constant but for larger fields it decreases at approximately the same rate as $T_1$. This reduction in $T_2$ is dominated by the decreasing energy difference between the AFD and the collinear AF phases with increasing magnetic field:
\begin{equation}
  T_2=[E_{AFD}^{\perp}(B)-E^\perp_{coll}(B)]/(S_{AFD}-S_{coll}).
\end{equation}
As the magnetic moments tilt in the direction of the magnetic field, the projection of the magnetic moments in the direction perpendicular to the field decreases, and it is this latter projection which determines the energy difference between the phases. The entropy difference between the two phases does not 
depend significantly on the external magnetic field.

\begin{figure}[t]
    \begin{center}
      \includegraphics[width=0.5\textwidth]{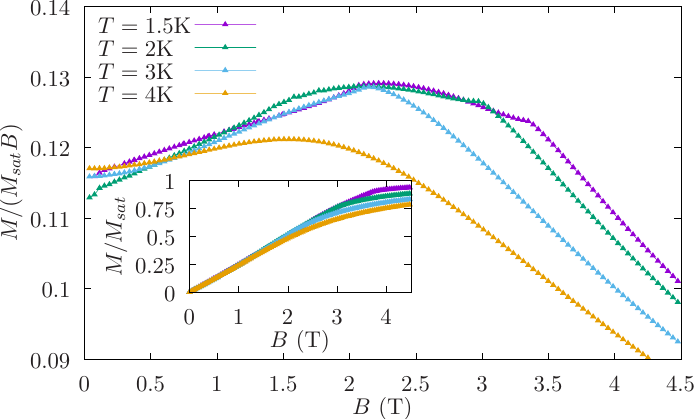}
    \end{center}
    \caption{(Color online) Magnetization as a function of the external magnetic field perpendicular to the easy axis $\hat{c}$ for different values of temperature. Other parameters as in Fig. \ref{fig:CvsT_B}.}
    \label{fig:Mag_perp}
\end{figure}

The magnetization $M$ as a function of the field intensity is shown 
in Fig. \ref{fig:Mag_perp}. Note that in contrast to the previous
case, the magnetization lies in the $a$--$b$ plane and 
does not show a jump at low temperatures, which is consistent with the absence of a spin-flop transition (see Fig. \ref{fig:Mag_perp}). At low temperatures, the magnetization divided by the magnetic field $M/B$ presents a maximum at the  AFD to collinear AF 
transition and a kink at the collinear to paramagnetic transition.  

\begin{figure}[t]
\begin{center}
       \includegraphics[width=0.45\textwidth]{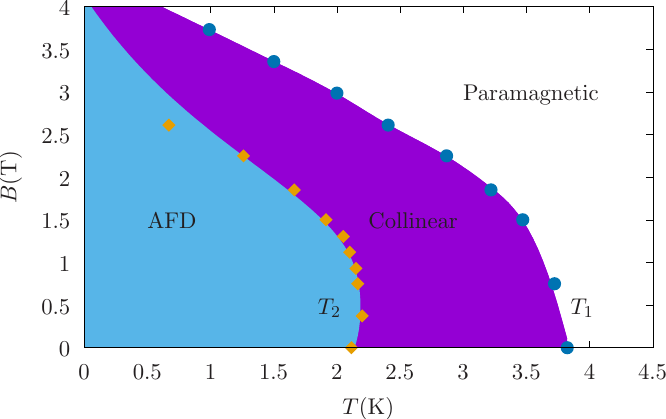}
    \end{center}
    \caption{Phase diagram for an external magnetic field perpendicular to the easy axis $\hat{c}$. Other parameters as in Fig. \ref{fig:CvsT_B}.}
    \label{fig:PDrot}
\end{figure}

The results for a magnetic field perpendicular to the easy axis are summarized in the phase diagram of Fig. \ref{fig:PDrot} \footnote{We were unable to determine precisely the behavior of the system at low but finite temperatures and intermediate fields perpendicular to the easy axis ($T<1$K and $B>5$T).}. For a non-zero magnetic field, the magnetic moments in the collinear phase are no longer along the $\hat{c}$-axis, but are tilted in the direction of the magnetic field.
The magnetic order in this phase can be qualitatively seen as the addition of a uniform magnetization, parallel to the external magnetic field, and a collinear antiferromagnetic order with the magnetic moments parallel to the $\hat{c}$-axis.

\subsection{Order parameter analysis of the magnetic phases}

To analyze the spin configurations in the different phases identified in Fig. \ref{fig:PD} we calculate the structure factor:
\begin{equation}
  \mathbf{S}(\mathbf{q})=\frac{1}{J N}\sum_{\ell}\mathcal{J}_\ell e^{ i \mathbf{q}\cdot \mathbf{R}_\ell},
  \label{eq:strfac}
\end{equation}
where $N$ is the number of \Eui\ ions, and $\mathbf{R}_\ell$ is the position of the magnetic moment $\mathcal{J}_\ell$.
We only obtain sizable values (i.e. $\mathcal{O}(1)$) of $|\mathbf{S}(\mathbf{q})|^2$, for $\mathbf{q}=0$ (provided $B\neq 0$), $\mathbf{q} =\mathbf{Q} \equiv 2\pi(\frac{1}{3a},\frac{1}{\sqrt{3}a},\frac{1}{2c})$, and symmetry related wave vectors. The wave-vector $\mathbf{Q}$ is consistent with the $a$-$b$ plane periodicity of the AF120 structure and with an antiferromagnetic order along the $\hat{c}$-axis. For a magnetic field parallel to the $\hat{c}$-axis we define (in keeping with the rotational symmetry of the system around the $\hat{c}$-axis) the order parameters: $s_\parallel=|\hat{c}\cdot \mathbf{S}(\mathbf{Q})|^2$ and $s_\perp=|\mathbf{S}(\mathbf{Q})|^2-s_\parallel$. We also define $m=|\hat{c}\cdot\mathbf{S}(0)|^2$ which is a measure of the degree of polarization of the magnetic moments along the $\hat{c}$-axis.
 \begin{figure}[t]
\begin{center}
       \includegraphics[width=0.45\textwidth]{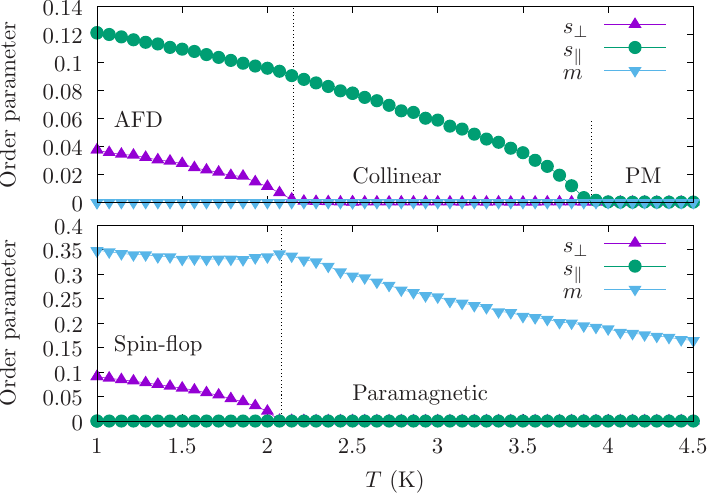}
    \end{center}
    \caption{(Color online) Order parameters (see text) as a function of the temperature at $B=0$ (top panel) and $B=1.5$T (lower panel). The other parameters are as in Fig. \ref{fig:CvsT_B}. The transition temperatures between the different phases are indicated with dotted style lines.}
       \label{fig:OP}
\end{figure}

Figure \ref{fig:OP} presents the order parameters as a function of the temperature for two values of the external magnetic field ($B=0$ and $\mathbf{B}=1.5$T$\hat{c}$).  
For $B=0$ we have no uniform magnetization ($m=0$) and at high temperatures, in the paramagnetic (PM) phase, there is no magnetic order: $s_\perp=s_\parallel=0$. 
At the transition temperature $T_1$ the magnetic moments order in a collinear structure which is characterized by $s_\parallel\neq 0$ and $s_\perp=0$. As the temperature is lowered further across $T_2$, the system enters the AFD phase in which both $s_\perp$ and $s_\parallel$ are finite. The emergence of a nonzero $s_\perp$ for $T<T_2$ can be interpreted as the ordering of the basal plane components of the magnetic moments. 

At low temperatures and $B=1.5$T the system is in the spin-flop phase which can be characterized by $s_\perp\neq 0$, $s_\parallel=0$, and $m\neq0$ (see lower panel in Fig. \ref{fig:OP}). The spin-flop state can be viewed as an AF120 state in the $a$-$b$ plane ($s_\perp\neq0$) with an added uniform magnetization along the $\hat{c}$-axis ($m\neq0$). Increasing the temperature in the spin-flop state leads to a decrase of $S_\perp$ which vanishes at the spin-flop to paramagnetic phase transition line.

The characterization of the different magnetic phases derived from order parameter analysis is consistent with the one deduced from the inspection of the different configurations of the magnetic moments and the analytical results.

\section{Summary and Conclusions}\label{sec:concl}
We present a detailed description of the magnetic properties of the chiral material \Eu. We propose a simplified model with exchange couplings up to next nearest neighbours to describe the interactions between the local magnetic moments on the \Eui\ ions. The parameters of the model are estimated using total energy calculations based on Density Functional Theory. 

We provide analytical expressions for the ground state energies of the different magnetic phases using a classical spin description for the $J=7/2$ magnetic moments.
We perform classical Monte Carlo simulations to calculate the specific heat and the magnetization as a function of the temperature for different values of the model parameters and the external magnetic field. We obtain a rich phase diagram including a generalization of the well known 120$^\circ$ structure in two dimensional systems. A collinear antiferromagnetic phase, stabilized at finite temperature by entropic effects, is also observed. This results in two phase transitions as a function of temperature, as observed experimentally \cite{franco2021synthesis}.

An external magnetic field parallel to the easy $\hat{c}$-axis reduces the temperature range of stability of the collinear phase leading to a single phase transition as a function of the temperature for high enough magnetic fields. For the model parameters analyzed in the numerical simulations, the suppression of the collinear AF phase is concomitant with a spin-flop transition.

We neglected in our analysis the Dzyaloshinskii-Moriya interaction.
While we do not expect the inclusion of this term to modify the essential features of the phase diagram,  it might stabilize a spiral order at low temperatures, possibly incommensurate, with a long wavelength 
in the $\hat{c}$ direction twisting the spin projections in the $a$--$b$ plane.

We expect this work to trigger further theoretical and experimental studies of this interesting material. 
\acknowledgements

We acknowledge financial support from grants PICT 2016/0204, 
PICT 2017-2726 and PICT 2018-01546 of the ANPCyT and
SeCTyP-UNCuyo grant 06/C569.

\appendix

\section{Ab initio calculation of the magnetic coupling parameters}\label{ap:couplings}
\begin{figure}[h]
  \centering
   \includegraphics[width=0.4\textwidth]{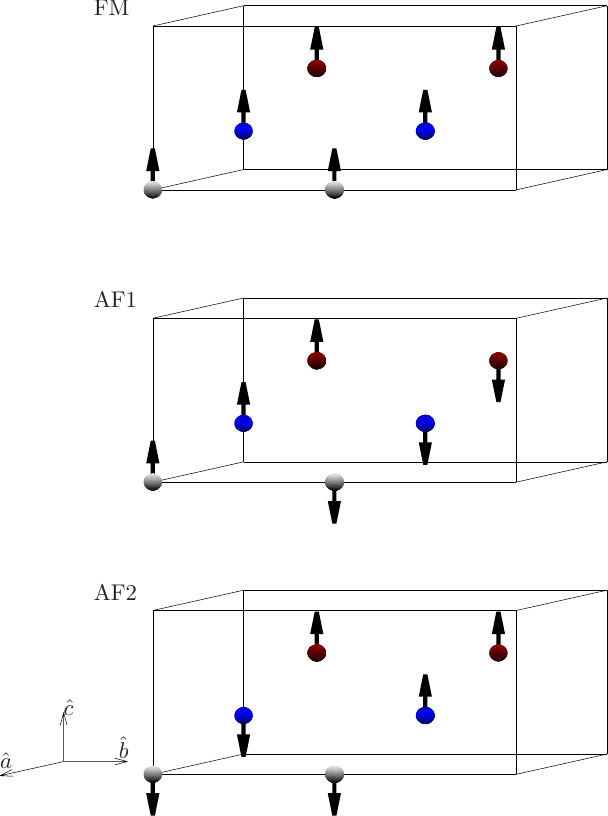}
   \caption{(Color online) Magnetic configurations evaluated for a $1\times2\times1$ cell (with periodic boundary conditions) to obtain the exchange coupling parameters. The orientation of the magnetic moments on the \Eui\ ions is indicated by black arrows. From top to bottom: ferromagnetic (FM), antiferromagnetic 1 (AF1), and antiferromagnetic 2 (AF2). The color coding is as in Fig. \ref{fig:Eucrystal}. } 
  \label{fig:configurations}
\end{figure}

We calculated the total energy for different static configurations of the local magnetic moments (see Fig. \ref{fig:configurations})\footnote{See Ref. \onlinecite{facio2015co} for a related calculation}. 

The total-energy calculations were performed using the generalized gradient approximation (GGA) of Perdew, Burke and Ernzerhof for the exchange and correlation functional as implemented in the Wien2K code \cite{wien2k,Perdew1996a}.
A local Coulomb repulsion was included in the Eu $4f$ shell and treated using GGA+U which is a reasonable approximation for these highly localized states. 
Due to the localized character of the $4f$ electrons, the fully localized limit was used for the double counting correction \cite{Anisimov1993}. 
We described the local Coulomb and exchange interactions with a single effective local repulsion $U_{eff} = U - J_H = 6 eV$ \cite{Yin2006,Petersen2006}.
The APW+local orbitals method of the \textsc{WIEN2K} code was used for the basis function \cite{wien2k}.

We consider the experimental lattice parameters reported in Ref. 
\onlinecite{lux1993kristallstrukturen}
($a$ = 6.671 \AA\ and $c$ = 7.055 \AA) and relaxed the internal positions. 
1000 $k$-points were used in the full Brillouin zone for the ionic relaxation in the unit cell, and 500 $k$-points for the $2\times1\times1$ supercell total-
energy calculations of the different magnetic configurations.

The total energy, relative to the ferromagnetic state, for each magnetic configuration and per \Eui\ ion is presented in Table \ref{tab:totnrg}.
\begin{table}
	\centering
    \begin{tabularx}{\columnwidth}{@{}l *2{>{\centering\arraybackslash}X}@{}}
		\hline
		\hline
		Order& energy difference with the FM state\\
		FM& 0\\
		AF1& $-2.7$ K	\\ 
		AF2& $1.95$ K\\ 
		 \hline
		\hline
	\end{tabularx}
	\caption{Relative energy $\Delta E$ (in Kelvin) per \Eui\ ion for the magnetic configurations of Fig. \ref{fig:configurations}}.
	\label{tab:totnrg}
\end{table}

In the absence of an applied magnetic field and using a classical magnetic moment description, the contribution per Eu atom to the total energy due to the magnetic interactions described in Eq. (\ref{eq:magham}) for the different configurations of Fig. \ref{fig:configurations} is given by:
\begin{equation}
    \begin{array}{r@{}l}
        E^m_{FM}/J^2 &={} -2 J_1 - J_2, \\ 
	E^m_{AF1} /J^2&={}\frac{1}{3} (-2 J_1 + J_2), \\
        E^m_{AF2}/J^2 &={}\frac{1}{3} (2 J_1 + J_2), \\
\end{array}
	\label{eq:couplings}
\end{equation}
where $J=7/2$ is the angular momentum of the Eu$^{2+}$ ion $4f$ electrons. The energy differences between magnetic configurations calculated from first principles can be combined with Eqs. (\ref{eq:couplings}) to obtain the coupling parameters $J_i$ solving a system of 2 linear equations. The results for the $J_i$ are presented in Table \ref{tab:exchcoup}.

\begin{table}
    \centering
    \begin{tabularx}{\columnwidth}{@{}l *2{>{\centering\arraybackslash}X}@{}}
        \hline
        \hline
	$J_1$& $0.28$ (FM)\\
	$J_2$&  $-0.45$ (AF)\\
        \hline
        \hline
    \end{tabularx}
    \caption{Calculated exchange couplings (in K)
    }
    \label{tab:exchcoup}
\end{table}

We also analyzed the crystal field induced magnetic anisotropy by performing total energy calculations including the spin-orbit coupling. We considered the AF1 magnetic moment configuration (see Fig. \ref{fig:configurations}) with all the magnetic moments parallel (or antiparallel) to the $\hat{a}$, $\hat{b}$ or $\hat{c}$ axis. The energies in Kelvin per \Eui\ ion are presented in Table \ref{tab:totnrgSO}.  The energy difference between the $\hat{a}$ and $\hat{b}$ orientations is of the order of the numerical error in the calculations. The $\hat{c}$ orientation has a significantly lower energy which is consistent with a $\sim 0.2$ K (easy axis) contribution to the anisotropy parameter $D$.

\begin{table}
	\centering
    \begin{tabularx}{\columnwidth}{@{}l *2{>{\centering\arraybackslash}X}@{}}
		\hline
		\hline
		magnetic moment orientation&	energy difference with the $\hat{c}$ orientation\\
		$\hat{a}$& 2.39 K\\
		$\hat{b}$& 2.46 K	\\ 
		$\hat{c}$& 0 \\ 
		 \hline
		\hline
	\end{tabularx}
	\caption{Relative energy $\Delta E$ (in Kelvin) per \Eui\ ion for the AF1 configuration (see Fig. \ref{fig:configurations}) and different orientations of the magnetic moments. }
	\label{tab:totnrgSO}
\end{table}

\section{Energy of the AFD state for a magnetic field parallel to the easy axis}\label{ap:AFDB}
\begin{figure}[h]
    \begin{center}
      \includegraphics[width=0.2\textwidth]{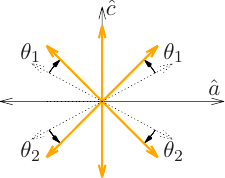}
    \end{center}
    \caption{(Color online) Schematic representation of the magnetic moment orientations under an external magnetic field parallel to the $\hat{c}$-axis for an easy axis anisotropy on the same axis.}
    \label{fig:anglesAFDB}
\end{figure}

An external magnetic field parallel to the easy axis ($\hat{c}$) breaks the mirror symmetry (about the $a$--$b$ plane) of the AFD state. Under a magnetic field, the shift angle of the magnetic moments is different depending on the sign of their projection along the $\hat{c}$-axis (see Fig. \ref{fig:anglesAFDB}).

\begin{figure}[h]
    \begin{center}
      \includegraphics[width=0.45\textwidth]{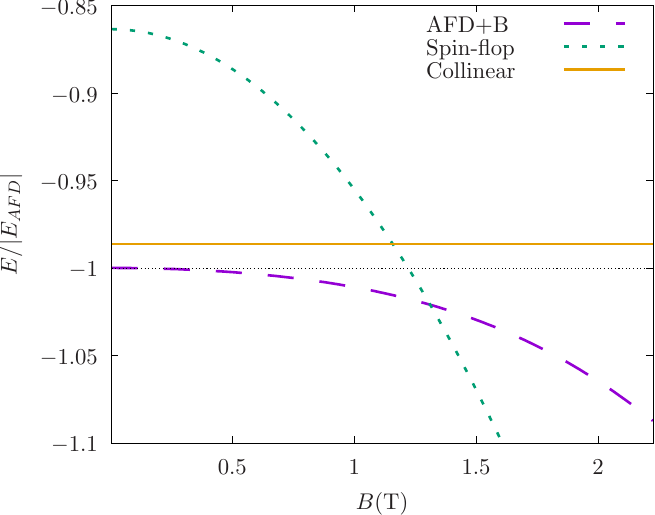}
    \end{center}
    \caption{(Color online)  Energy of the AFD+B (the AFD state distorted by the magnetic field), spin-flop and collinear antiferromagnet states as a function of the magnetic field. Other parameters as in Fig. \ref{fig:CvsT_B}.}
    \label{fig:Eflop}
\end{figure}

The energy as a function of the two possible shift angles $\delta\theta_1$ and $\delta\theta_2$ (see Fig. \ref{fig:anglesAFDB}) reads: 

\begin{align*}
  E=&-\tfrac{1}{3} B J \left(\cos\left(\tfrac{\pi }{3}-\theta_1\right)-\cos
   \left(\tfrac{\pi }{3}-\theta_2\right)\right)\\
&  -\tfrac{1}{3} D J^2 \left(\cos^2\left(\tfrac{\pi }{3}-\theta_1\right)+\cos^2\left(\tfrac{\pi }{3}-\theta_2\right)+1\right)\\
   &-\tfrac{2}{3} J^2 J_1 \left[\cos
   \left(\theta_1+\theta_2+\tfrac{\pi }{3}\right)+ \cos \left(\tfrac{\pi }{3}-\theta_1\right)\right.\\
   &\left. +\cos \left(\tfrac{\pi}{3}-\theta_2\right)\right]\\
   &+\tfrac{1}{3} J^2 J_2 (2 \cos (\theta_1-\theta_2)+1).
\end{align*}
We minimized this energy w.r.t  $\delta\theta_1$ and $\delta\theta_2$ for different intensities of the external magnetic field. The results are presented in Fig. \ref{fig:Eflop} and compared to the spin-flop state energy which allows to determine the spin-flop field. For the values of the Hamiltonian parameters considered, the magnetic field that produces the spin-flop transition differs less that 10\% for the different orders considered (collinear and AFD). We approximate the spin-flop field for the AFD state (see Eq. \ref{eq:flopafd}) as the one that makes the energy of the spin-flop state (see Eq. \ref{eq:eflop}) equal to $E_{AFD}$ (see Eq. \ref{eq:EAFD}), i.e. neglecting the shift in the angles produced by the external magnetic field.

\bibliography{115b,chiral,triang}{}
\bibliographystyle{apsrev4-1}

\end{document}